\newif\ifisanon
\isanonfalse

\ifisanon
  \documentclass[10pt,sigconf,letterpaper,anonymous]{acmart}
\else
  \documentclass[10pt,sigconf,letterpaper]{acmart}
\fi

\usepackage{tikz}
\usepackage{amsmath}

\usepackage{multirow}
\usepackage{url}
\usepackage{multicol}
\usepackage{subcaption}
\usepackage{graphicx} 
\usepackage{hyperref}
\usepackage{amsmath}

\usepackage{xspace}

\usepackage{tikz} 
\usetikzlibrary{arrows,automata}
\usepackage[title]{appendix}

\hypersetup{colorlinks,linkcolor=[rgb]{.75,0,0},urlcolor=[rgb]{0.75,0,0.75},citecolor=blue,breaklinks} 

\makeatletter
  \newcommand\EatSpacesHack{\@bsphack\@esphack}
\makeatother

\iffalse
  \renewcommand\comment[1]{{\color{blue} \sffamily [xxx:  #1]}}
  \newcommand\commentceron[1]{{\color{red} \sffamily [xxx:  #1 - João Ceron - \today]}}
  \newcommand\leandro[1]{{\color{orange} \sffamily [xxx:  #1 - Leandro - \today]}}
  \newcommand\PostSubmission[1]{\EatSpacesHack}
  \newcommand\reviewfix[1]{{\sffamily [RF:#1]}\@bsphack\@esphack}
\else
  \renewcommand\comment[1]{\EatSpacesHack}
  \newcommand\commentceron[1]{\EatSpacesHack}
  \newcommand\leandro[1]{\EatSpacesHack}
  \newcommand\reviewfix[1]{\EatSpacesHack}
  \newcommand\PostSubmission[1]{\EatSpacesHack}
\fi
\makeatother

\def\Snospace~{\S{}} %
\date{}

\ifisanon

\newcommand\anonymize[1]{\textit{(anonymized for review)}}
\else

\fi

\begin{document}

\title{Chhoyhopper: A Moving Target Defense with IPv6}

\ifisanon
  \author[anonymized]{Authors Anonymized for Review}
\else
 \author[Rizvi and Heidemann]{
   {ASM Rizvi and  John Heidemann }}
  \affiliation {USC/Information Sciences Institute and USC/CS Dept. }
\fi

\begin{abstract}
Services on the public Internet are frequently scanned,
  then subject to brute-force and denial-of-service attacks.
We would like to run such services stealthily,
  available to friends but hidden from adversaries.
In this work, we propose a moving target defense
  named ``Chhoyhopper'' that utilizes the
  vast IPv6 address space to conceal publicly available services.
The client and server 
  to hop to different IPv6 addresses in a pattern based on a 
  shared, pre-distributed secret and the time-of-day.
By hopping over a /64 prefix,
  services cannot be found by active scanners,
  and passively observed information is useless after two minutes.
We demonstrate our system with SSH,
  and show that it can be extended to other applications.
\end{abstract}

\keywords{IPv6, Moving target defense, SSH}

\maketitle

\pagestyle{plain}


\section{Introduction}
        \label{sec:intro}

Services on the public Internet are frequently scanned,
  then subject to brute-force and denial-of-service attacks.
IPv4 scanning has been possible for more than a decade~\cite{Heidemann08c}
  and recent tools allow scanning all of IPv4 in minutes~\cite{Adrian14a,Graham14a}
Regular scanning is done by many parties~\cite{Wustrow10a}

We would like to provide \emph{stealthy} services on the public Internet,
  available to friends but hidden from adversaries.


IPv6 provides a huge address space in which we can hide services.
In spite of attempts to discover active addresses,
  when every LAN has $2^{64}$ addresses (or more),
  active discovery of services on intentially obscure addresses
  is intractable (see \autoref{sec:discovery}).
With /48s as the recommended minimum size of publically
  routable prefix~\cite{apnic-64prefix},
  and /56s recommended for homes~\cite{Narten11a},
  even with a million devices in a home,
  quintillions of addresses remain unused on every network.
  
The contribution of our paper is to propose a moving target defense,
  \emph{Chhoyhopper}\footnote{Chhoy is the number ``six'' in Bengali, since we hop in IPv6.}, that uses the
  vast IPv6 address space to conceal publicly available services.
The client and server 
  to hop to different IPv6 addresses in a pattern based on a 
  shared, pre-distributed secret and the time-of-day.
By hopping over a /64 prefix,
  services cannot be found by active scanners,
  and passively observed information is useless after two minutes.
We demonstrate our system with SSH,
  and show that it can be extended to other applications.

\textbf{Related work:}
Our work builds on ideas in privacy-preserving IPv6 address assignment~\cite{gont2014method,Gont21a},
  but while that work proposes updating addresses daily with a fixed pattern,
  we accelerate hopping each minute to service as an active defense against scanning.
Our work is similar to port knocking~\cite{Krzywinski03a,degraaf2005improved},
  but it hides in IPv6 rather than requiring ``wake-up'' packets.
Closest to our work is IPv4-based port-hopping~\cite{lee2004port};
  we take advantage of much larger IPv6 space ($2^{64}$) compared to the
  quite limited IPv4 port space ($2^{16}$).

\textbf{Availability:}
Our implementation is freely available at \url{https://ant.isi.edu/software/chhoyhopper/}.

\section{Chhoyhopper Design}
                \label{sec:system}

\begin{figure}
        \centering
        \includegraphics[width=0.9\linewidth]{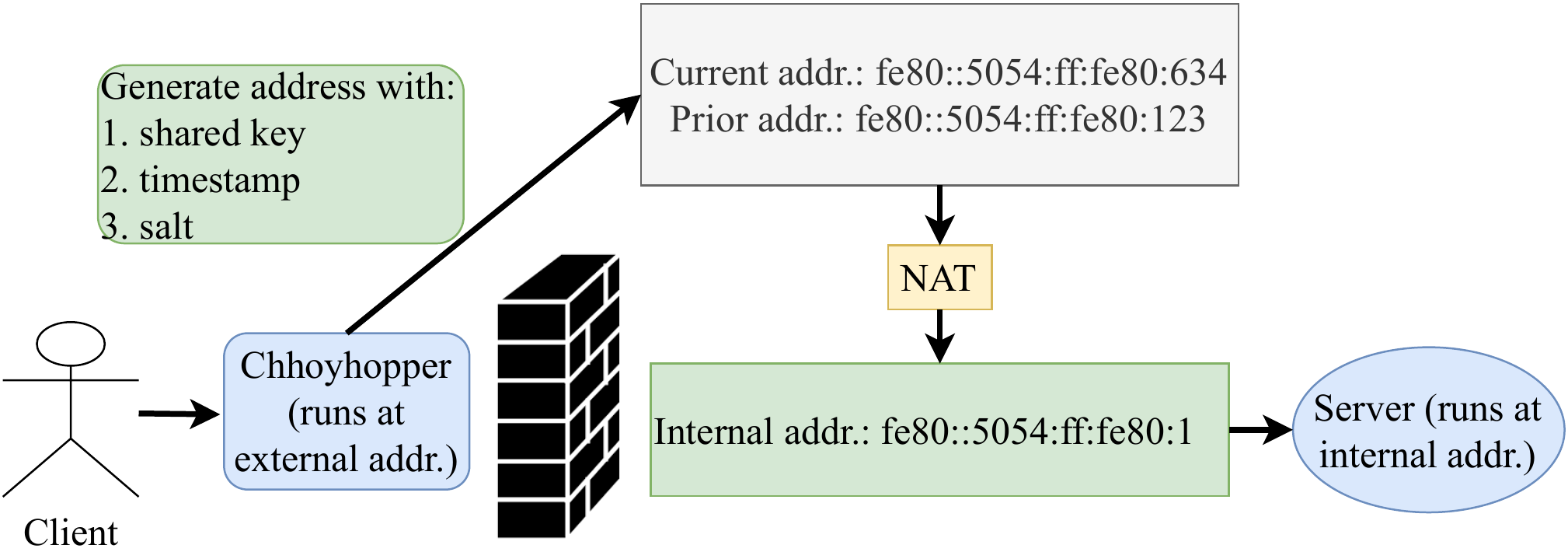}
        \caption{Client and server interaction in Chhoyhopper}
        \label{fig:chhoy-hopper-interaction}
\end{figure}

Our goal is to allow the client to rendezvous with the server
  on a public, but temporary IPv6 address.
By allocating the temporary address from a large space ($2^{64}$ addresses),
  scanning is impractical, as we show in \autoref{sec:discovery}.
By changing the address frequently,
  reuse of a passively observed temporary address is only possible
  for a very brief window of time.
The hopping pattern is cryptographically secure,
  so prior active addresses reveal nothing about future addresses.

\autoref{fig:chhoy-hopper-interaction} shows the components of our system,
  and we describe them next:
  selection and lifetime of the temporary address,
  hopping on the server,
and hopping by the client.

\textbf{Address Hopping Pattern:}
The client and server must follow the same hopping pattern to rendezvous.
We assume they share a pre-distributed
  secret key, which may be distributed by
  several means, such as
  face-to-face sharing ahead-of-time,
  through a secure channel such as encrypted e-mail.
Our requirement for this secret means Chhoyhopper cannot be used
  for anonymous clients to discover a server,
  since scanners could exploit any discovery process.

The server and the client compute the same temporary address
  by computing a cryptographic hash of the shared secret,
   a salt value, and the current time in minutes.
We use the SHA-256 algorithm for hashing
  and the time in seconds since the Unix epoch.
The salt value prevents rainbow attacks~\cite{Oechslin03a}
  and can vary by service or deployment.

We compute the IPv6 address in two parts.
We take the DNS name of the service address
  and look up a full IPv6 address,
  but replace the low 64-bits of the address with the top 64-bits of the
  hash result.
Use of DNS allows the service to move in the Internet
  and provides a user-friendly name.
DNSSEC should be used to ensure that the DNS lookup
  of the top IPv6 address bits is not subject to a person-in-the-middle attack.
We discuss the potential of collisions in \autoref{sec:collisions}.

\textbf{Server-Side Hopping:}
The server tracks its current address, changing it every minute.
To avoid problems with clock skew, the server listens to \emph{two} addresses,
  one for the current minute and the other for the nearest adjacent minute.

It is cumbersome
  for server software to change its service address every minute,
  and we would rather not modify server software.
We therefore operate the server on a fixed address
  that is firewalled from the public Internet.
A daemon then uses network address translation
  to map the currently active addresses
  through the firewall to the internal fixed address.
NAT translation also ensures that once a connection is established
  it continues to operate, even after the server moves to other
  addresses for new connections.

\comment{proposed new para.  Note that my i/ii/iii are differnet than yours. ---johnh 2021-05-04 This looks ok to me. I am commenting out the previous para. --- asmrizvi 2021-05-05}
To summarize server processing in \autoref{fig:chhoy-hopper-interaction}:
  (i) new flows to the current and prior address are detected by NAT rules
    and establish new connection state before being passed to the internal server address,
  (ii) existing flows are detected by NAT and pass through to the internal address.
(iii) Any other addresses, including external traffic sent to the ``internal''
  server address, are dropped by the server's firewall.

Our NAT-manipulation daemon is a simple Python program
  modifying Linux ip6tables.
The daemon assigns the NAT rules to a particular external interface on the server.
\comment{does this mean we don't work on multi-homed servers? ---johnh 2021-05-04 I haven't tested that. But my assumption is our system should still work. We should be able to use different internal addresses. Or we can try to forward the packets to a different interface with our internal address. ---asmrizvi 2021-05-05}

\textbf{Hopping at the Client:}
The client must compute and use 
  the server's current IPv6 address to begin a new connection.
We assume the server's secret key and the salt are known to the client,
  so the client does the same hash computation as the server.
As with the server, the client looks up an IPv6 address from DNS 
  and replaces the low-64 bits with the current temporary hash.

Our client implementation for SSH uses a simple Python program
  which invokes the native client with appropriate arguments.
  \comment{is next sen true? ---johnh 2021-05-04 I hope so. Need to start working on it. ---asmrizvi 2021-05-05}
We also plan to provide a Chhoyhopper client
  as a patch to OpenSSH.

\textbf{Other Applications:}
\comment{new paras, please review ---johnh 2021-05-04}
To date we have only implemented Chhoyhopper for SSH.
In principle it can apply to any connection-based application.
We have considered, but not yet implemented, an implementation
  for HTTP\@.

The main challenge in supporting an application is to
  transparently interpose between the client and the server.
Since our implementation requires no server-side changes,
  supporting new servers is easy.

Our client implementation for SSH requires the users to employ a new
  SSH front-end when they start the connection.
A client for HTTP is more difficult since a web browser regularly
  creates many new connections.
We have considered two approaches:
  a browser-side plugin could detect and rewrite outgoing connections
  to hosts that match servers with Chhoyhopper support.
Second, an HTTP proxy could handle this mapping.
Both approaches are potential future work.

\section{Analysis}

\textbf{Risk of Discovery:}
       \label{sec:discovery}
To estimate the difficulty of brute-force scanning,
  consider a scanner scanning at 100Gb/s
  looking for a server hopping in one /64 with 64B TCP SYNs.
%
%
At that rate (scanning $2\times10^{8}$ addresses per second)
  the expected time to discover one server is about 3000 years,
  at which point the adversary will have at most two minutes to exploit it.
Since the address space is huge compared to the scanning rate,
  we are confident that brute-force scanning is impractical.
Since the address is hopping randomly, intelligent scanning is not possible.

\textbf{Risk of Collisions:}
        \label{sec:collisions}
When multiple servers share the same /64 address prefix,
  it is possible that they could collide and hop to the same address.
\comment{I've changed the following line. ---asmrizvi 2021-05-05}
A concerned operator should assign a unique IPv6 address
  every minute that is not used by any other server.
However, we suggest that odds of collision is so low
  that collision avoidance is unnecessary.


Collisions of hopping addresses is equivalent to the
  well-known Birthday Problem,
  but rather than $n$ people in 365 days of the year,
  we have $k$ servers in $2^{64}$ addresses.
Using a simplified approximation,
  the probability of a hash collision in any given minute
  is $1 - e^{\frac{-k(k - 1)}{2N}}$~\cite{hash-collision}.
Using this formula,
  the probability of an address
  mapped into the $k$ of 1\,million addresses is only
  1 in 37\,million.
As we generate an address every minute,
  we can expect a collision with these million servers
  once in every 70 years.
This failure rate is considerably less than
  DRAM failures due to cosmic radiation~\cite{Sridharan12a}.

\section{Conclusions}
In this paper,
  we provide an implementation of a moving target defense
  named ``Chhoyhopper'' to provide security utilizing
  the huge IPv6 address space.
Using our system, a service will hop over
  different IPv6 addresses, and a client needs to
  find the current IPv6 address to connect.
We implement our approach for SSH application,
  and in future we plan to provide support for
  other applications.

\ifisanon
\else
\section*{Acknowledgments}

ASM Rizvi and John Heidemann's work on this paper is supported, in part, by 
 the DHS HSARPA Cyber Security Division via contract number HSHQDC-17-R-B0004-TTA.02-0006-I,  
  and by DARPA under Contract No.~HR001120C0157.  
Any opinions, findings and conclusions or recommendations expressed
  in this material are those of the authors
  and do not necessarily reflect the views of NSF or DARPA.

We thank Rayner Pais who prototyped an early version of Chhoyhopper and an IPv4 port hopper.

\fi

\bibliographystyle{plain} 
\bibliography{paper} 

\begin{thebibliography}{10}

\bibitem{Adrian14a}
David Adrian, Zakir Durumeric, Gulshan Singh, and J.~Alex Halderman.
\newblock Zippier {ZMap}: {Internet}-wide scanning at 10 {Gbps}.
\newblock In {\em Proceedings of the USENIX Workshop on Offensive
  Technologies}, San Diego, CA, USA, August 2014. {USENIX}.

\bibitem{degraaf2005improved}
Rennie Degraaf, John Aycock, and Michael Jacobson.
\newblock Improved port knocking with strong authentication.
\newblock In {\em 21st Annual Computer Security Applications Conference
  (ACSAC'05)}, pages 10--pp. IEEE, 2005.

\bibitem{gont2014method}
F.~Gont.
\newblock A method for generating semantically opaque interface identifiers
  with {IPv6} stateless address autoconfiguration ({SLAAC}).
\newblock RFC 7217, Internet Request For Comments, April 2014.

\bibitem{Gont21a}
F.~Gont, S.~Krishnan, T.~Narten, and R.~Draves.
\newblock Temporary address extensions for stateless address autoconfiguration
  in ipv6.
\newblock RFC 8981, Internet Request For Comments, February 2021.

\bibitem{Graham14a}
Robert Graham, Paul McMillan, and Dan Tentler.
\newblock Mass scanning the internet.
\newblock Presentation at Defcon 22, August 2014.

\bibitem{Heidemann08c}
John Heidemann, Yuri Pradkin, Ramesh Govindan, Christos Papadopoulos, Genevieve
  Bartlett, and Joseph Bannister.
\newblock Census and survey of the visible {Internet}.
\newblock In {\em Proceedings of the ACM Internet Measurement Conference},
  pages 169--182, Vouliagmeni, Greece, October 2008. {ACM}.

\bibitem{Krzywinski03a}
Martin Krzywinski.
\newblock Port knocking: Network authentication across closed ports.
\newblock {\em SysAdmin Magazine}, 12(6):12--17, June 2003.

\bibitem{lee2004port}
Henry~CJ Lee and Vrizlynn~LL Thing.
\newblock Port hopping for resilient networks.
\newblock In {\em IEEE 60th Vehicular Technology Conference, 2004.
  VTC2004-Fall. 2004}, volume~5, pages 3291--3295. IEEE, 2004.

\bibitem{Narten11a}
T.~Narten, G.~Huston, and L.~Roberts.
\newblock {IPv6} address assignment to end sites.
\newblock RFC 6177, Internet Request For Comments, March 2011.

\bibitem{Oechslin03a}
Philippe Oechslin.
\newblock Making a faster cryptanalytic time-memory trade-off.
\newblock In {\em Proceedings of the {IACR} CRYPTO}, volume 2729, pages
  617--630. International Association for Cryptologic Research, August 2003.

\bibitem{hash-collision}
Preshing on~Programming.
\newblock Hash collision probabilities.
\newblock \url{https://preshing.com/20110504/hash-collision-probabilities/},
  2011.
\newblock [Online; accessed 15-March-2021].

\bibitem{Sridharan12a}
Vilas Sridharan and Dean Liberty.
\newblock A study of {DRAM} failures in the field.
\newblock In {\em Proceedings of the {ACM} SuperComputing}, pages 1--11, Salt
  Lake City, Utah, USA, November 2012. {ACM}.

\bibitem{apnic-64prefix}
Jessica Wei.
\newblock Why is a /48 the recommended minimum prefix size for routing?
\newblock
  \url{https://blog.apnic.net/2020/06/01/why-is-a-48-the-recommended-minimum-prefix-size-for-routing/},
  2020.
\newblock [Online; accessed 15-March-2021].

\bibitem{Wustrow10a}
Eric Wustrow, Manish Karir, Michael Bailey, Farnam Jahanian, and Geoff Houston.
\newblock {Internet} background radiation revisited.
\newblock In {\em Proceedings of the 10th ACM Internet Measurement Conference},
  pages 62--73, Melbourne, Australia, November 2010. {ACM}.

\end{thebibliography}



\end{document}